\documentclass[12pt]{article}
\usepackage{graphicx}
\newcommand{\be}{\begin{equation}}
\newcommand{\bea}{\begin{eqnarray} \nonumber}
\newcommand{\ee}{\end{equation}}
\newcommand{\eea}{\end{eqnarray}}

 \def\(({\left(}
 \def\)){\right)}
\def\[[{\left[}
\def\]]{\right]}

\def \form#1 {eq. (\ref{#1}) }
\def \parziale#1#2  {{\partial {#1} \over \partial {#2}}}

\def \ba#1 {\overline{#1}}

\begin{document}

\title{On local equilibrium equations for clustering states}
\author{ Giorgio Parisi\\
Dipartimento di Fisica, Sezione INFN, SMC and
 UdRm1 of INFM,\\
Universit\`a di Roma ``La Sapienza'', \\
Piazzale Aldo Moro 2,
I-00185 Rome (Italy)}

\maketitle

%{Large}
\begin{abstract}

\noindent In this note we show that local equilibrium equations (the generalization of the TAP equations or of the 
belief propagation equations) do have solutions in the colorable phase of the coloring problem.  The same results extend 
to other optimization problems where the solutions has cost zero (e.g. K-satisfiability).  On a random graph the 
solutions of the local equilibrium equations are associated to clusters of configurations (clustering states).  On a 
random graph the local equilibrium equations have solutions almost everywhere in the uncolored phase; in this case we 
have to introduce the concept quasi-solution of the local equilibrium equations.
 
\end{abstract}

\section{Introduction}

Statistical mechanics of complex systems, in the limit where the number $N$ of degrees of freedom goes to infinity, has been
widely investigated in these recent years \cite{MPV}. A crucial  role is played by the local equilibrium states: often  a system is 
declared complex only  if the the number of local equilibrium states is exponentially large. Intuitively we would say 
that there are many local equilibrium states if the configuration space breaks down into many unconnected (or 
connected through channels of low probability) clusters of configurations.  The breaking of the configuration space 
into an exponential number of clusters corresponds to the breaking of the replica symmetry and it is one of the crucial 
 phenomena described by the replica approach \cite{MPV,LH}.

It should be possible to make a precise mathematical statement that corresponds to the intuitive notion of having an exponentially 
large number of states in the limit $N$ going to infinity.  Unfortunately a mathematical precise definition of a local 
equilibrium state for a given system with large $N$ is not easy, especially when the number of a local equilibrium states 
is very large.  A tentative definition of a local equilibrium state has been done using dynamical ideas \cite{BM}.  One introduce 
a \emph{natural} dependence of the configurations on the time and a local equilibrium state is a region where the system 
may remain for an exponentially large time.

This definition may be adequate for some purposes, but it is not so natural as far as it can be done only after the 
introduction of a dynamics.  Often one try to associate local equilibrium states to solutions of the local equilibrium 
equations.

The local equilibrium  equations are called in the literature under different names in different contexts: in statistical 
mechanics they are the TAP equations\ \cite{TAP}, in optimization problems the finite temperature TAP equations correspond to the 
`Sum-Product'' algorithm and the zero temperature limit of the TAP equations are the belief propagation equations that 
correspond to ``Min-Sum'' algorithm \cite{BP,factor}.

One difficulty with the approach based on the local equilibrium equations is that at the present moment we do not have a 
clear idea if these equations do have a solution and, if solutions do exist, how are they related to local equilibrium 
states.

The aim of this note is to get a better understanding of these matters; this will be done by presenting explicit solutions 
(or quasi solutions) of the local equilibrium equations at zero temperature.  This will be done for problems defined on a
random graph with a finite number of edges starting from each node.  The 
situation in infinite range models, like the Sherrington Kirkpatrick model, is likely different.
    
These ideas have been  used to study problems defined on  random graphs. The prototype of a random graph we will consider 
in this paper is a randomly chosen graph among all possible graphs with $N$ nodes and $M=\alpha N$ links.

The same considerations can be done also for other problems, like the k-satisfiability, where similar computations have 
been done.
\section{Coloring a colorable graph}

\subsection{Whitening a colored graph}

Let us consider a graph with $N$ nodes and $M=\alpha N$ links. Our task is to color it by assigning a color to each 
node in such a way that two adjacent nodes have different color. The number of colors is fixed and it is equal to $q$.

The interesting 
situation is where the graph is colorable in many different way (i.e. much large of $q!$). If the graph is colorable, we 
are interested in knowing its entropy density $S$ ($NS$ is the total entropy) defined by
\be
\#(coloring)= \exp(N S).
\ee

For a colorable graph our aim is to associate to each coloring a solution of the local equilibrium equations, that will 
be defined later.  We proceed in the following way.  A coloring is a function $c(i)$, where $i$ is a node and $c(i)$ can 
takes integer values in the interval $[1-q]$.  The coloring is legal if $c(i)\ne c(j)$ for all nodes adjacent to $i$.

Starting from a legal coloring we can introduce a whitening procedure.  We examine each node.  We have two 
possibilities:
\begin{itemize} 
    \item The color of the node is blocked and it cannot be changed producing a legal coloring: this happens if  the 
    node has neighbours with exactly $q-1$ different (non-white) colors.
    \item
    The color of the node is \emph{not} blocked and it can take at least two different legal colors: this happens if  the 
    node has neighbours with \emph{less} than $q-1$ different (non-white) colors.
\end{itemize}
We assign to the node $i$ the white color (conventionally 0) if we can produce a legal coloring by changing the color of 
the node $i$ without changing the color of the adjacent nodes (i.e. in the second case).  This procedure is repeated by 
considering  the non-white adjacent nodes until no nodes can be whitened \footnote{Similar ideas are common in the 
literature: sometimes white node are called \emph{spin fou.}. In the approach of \cite{MP1,MP2} their cavity fields are 
zero. They play an important role and they are called jokers in \cite{BMZ2}).}.

In other words a coloring is legal if
\be
\delta_{c(i),c(j)}=0 \ \ \ \forall\ \  j\in A(i) \ ,
\ee
where $A(i)$ is the set of nodes adjacent to $i$.  A whitening $w(i)$  can takes integer values in the interval 
$[0-q]$; it is legal if
\be
w(i) w(j)\delta_{w(i),w(j)}=0 \ \ \ \forall\ \  j\in A(i) \ .
\ee

A whitening is extremal if it cannot be whitened.  In this way we associate to each coloring an extremal whitening.   It 
is crucial that at each step the number of white nodes increases, so that the whitening procedure must stop somewhere.

It is possible to prove that the extremal whitening associated to a coloring is unique, i.e. it does not depend on the 
order in which we perform the whitening.  Extremal whitenings satisfy simple local equations: a node is white if the 
adjacent nodes have less than $q-1$ different non-white colors; a node is of the appropriate color if the adjacent nodes 
have exactly $q-1$ different non-white colors.

It is clear that sometimes the whitening produces a configuration where all nodes are white while in other cases we end 
in configurations in which some nodes are colored.  The whitening process is supposed to evidentiate those nodes that 
are essential in the coloring (i.e. the nodes that remain non-white); this extends the notion of core introduced in 
\cite{MRZ}.  We expect that in many cases a very large number of different legal coloring ends in the same extremal 
whitening.

There are many questions that can be posed. 
\begin{itemize}
    \item
If an extremal whitening is obtained by whitening a legal coloring, how difficult is the reconstruction of a legal coloring 
starting from the whitening? Intuitively is should be easier that starting from scratch, but it is not easy to quantify 
it.
\item
Are there solutions to the local equilibrium equations  for an extremal whitening that cannot be obtained by whitening a legal coloring?
\end{itemize}

The rational for introducing the whitenings is to extract some common properties from a some wide class of coloring (it 
  extends the notion of core introduced in \cite{MRZ}.  It 
is easy to argue that in the general case there is an exponentially large number of different legal coloring that 
produce the same extremal whitening.

\subsection{Directional coloring and directional whitening}

A directional coloring is a function $c(i|k)$, where $k \in A(i)$, that takes values in the interval $[1-q]$. It is a 
legal directional coloring if
\be
\delta_{c(i|k),c(j|m)}=0 \ \ \ \forall\ \  j\in A(i) \ \ \mbox{and} \ \  j \ne k\ 
\ \ \ \forall\ \  m\in A(j) \ \ \mbox{and} \ \  m \ne i\ .
\ee
In some sense  $c(i|k)$ is a color that the node $i$ could have in absence of the adjacent node $k$.

If we set 
\be
c(i|k) = c(i)
\ee
and $c(i)$ is a legal coloring,  $c(i|k)$ is a legal directional coloring. In the same way a directional whitening is a similar 
function $w(i|k)$ that takes values in the interval $[0-q]$. It is a legal  whitening if 
\be
w(i|k)w(j|i)\delta_{w(i|k),w(j|m)}=0, \ \ \ \forall\ \  j\in A(i) \ \ \mbox{and} \ \  j \ne k\
\ \ \ \forall\ \  m\in A(j) \ \ \mbox{and} \ \  m \ne i\ .
\ee

We can now start from a legal coloring, we produce a legal  directional coloring and we begin a whitening procedure where a 
$w(i|k)$ is set to zero if we can produce an other  legal directional whitening by changing the color $w(i|k)$ without
changing the color of the adjacent nodes. The procedure goes on up to the point where we reach an extremal directional 
whitening. 

Also in this case an extremal directional whitening can be characterized by local equations: it a legal whitening and it 
cannot be whitened; $w(i|k)$ is not white if and only if  there are exactly $q-1$ different non-white colors in the set 
formed by $w(j|m)$ with $j\in A(i) $, $m\in A(j) $, $j\ne k$  and $j \ne m$.  The local equilibrium conditions for a directional 
whitening to be extremal are equivalent to the TAP equations and to the belief equations in this context.

It is possible to prove that this construction is unique and there is one and only one extremal directional whitening 
associated  to a given legal coloring.  For such an extremal directional whitening,  for a given $i$ all non-white $w(i|k)$ must have 
the same color. Also if  this condition is satisfied it is not evident if there are  of extremal directional whitening that cannot 
be obtained by whitening a legal coloring.

Up to now we have only given definitions. The extremal whitening are useful because of the following three theorems.

(A)  If the graph is a tree and if we fix the value of the whitening on the leaves of the tree there is only one 
extremal whitening on the rest of the tree.

(B) If we consider two different legal colorings and the nodes where the two colorings  differ are contained in the interior of a tree 
(that is a subgraph of the original graph), the extremal directional whitening corresponding to these two coloring do coincide.
The same results hold if the nodes where they differ are contained in the interior of some non-overlapping  trees.

Theorem (A) is a particular case of the well known theorem on the convergence of the belief algorithm on a tree 
 \cite{factor} (an elementary proof for K-satisfiability can be found in \cite{BMZ}).

The proof of the theorem (B) is trivial. We start the whitening procedure in the interior of the tree. According to 
theorem (A) the whitening procedure will converge to a result that does not depend on the initial conditions. The 
difference among the different coloring is thus wiped up. We can now whitening the rest of the graph and the results are 
obviously equal in both cases. 

If we consider a random graph and a node $i$ of the graph. Let us call $Q(N,\alpha,L)$ the probability that two colorings 
that differ on some nodes, all of them at a distance less than $L$ from the node $i$, produce  two different extremal 
directional whitenings. We can prove the following theorem:

(C) The probability $Q(N,\alpha,L)$ goes to zero when $N$ goes to infinity at fixed $\alpha$.

Theorem (C) is a trivial consequence of theorem (B) and of the fact that for a random graph, when $N$ goes to infinity 
at fixed $\alpha$, the set of nodes at distance less than $L$ from a given node belongs to a tree with probability 1. In 
this situation a random graph is locally a tree.

Theorem (C) implies that when $N$ goes to infinity in a random graph two colorings must be different in a large number of 
nodes in order to produce  two different extremal directional whitenings. In other words, with probability one, colourings 
that differ only in a fixed amount of nodes do produce the same extremal directional whitening.

The local equilibrium equations for an extremal directional whitening are the zero temperature limit of the TAP 
equations for the coloring problem; it is believed that on a random graph we associate a different local equilibrium 
state to each different solution of such equations .

\section{Coloring uncolorable graphs}
\subsection{Directional whitening does not work}
In the case where the graph if not colorable we can  ask how to colour it in such a way to minimize  the number of edges 
of different color. 
In this case we would like to find the 
minimum of the quantity
\be
H=\sum_{i,j; j\in A(i)} \delta_{c(i),c(j))} \ .
\ee

We will say that a given configuration is a local minimum of order $k$  (or $k$-stable) if the energy cannot decrease if we change the 
color in a set of nodes that are located at distance equal or less than $k$ from a given node \cite{BM}.  If $k$ is large enough 
the local minimum coincides with the global minimum.

We are interested to study the case of a random graph when $N$ goes to infinite at fixed $\alpha$ in the regime where
\be
1<<k<<N \ .
\ee

We could firstly try to use the same approach of the previous section.

The most naive approach would be to start from the actual coloring and to construct a directional whitening by choosing 
the color of the node $i$ in absence of the edge from $i$ to $j$ by knowing the colors of the others nodes near to $i$.
If more that one color is possible, we can set the color to white (in certain cases we should also remember the list of the 
allowed colors for a non-white node). 

However here in this process a node can go from a non-white color to an other non-white color (in one or two steps) and the 
iterative process may not converge \cite{MPZ,CGPM}.  

There is a very simple case that displays this behaviour.  We consider a ring with 
$N$ nodes, $N$ being an \emph{odd} number: the simplest case is $N=3$ and $q=2$.  The iterative equations are
\be
c(i,i+1)=f(c(i-i,i)) \label{TAP.EQ}
\ee
where the sums are done modulo $N$. The function $f$ is defined by $f(1)=2$ and $f(2)=1$. 

It is evident that the previous equations have no solution and, if we iterate then, we find an oscillatory sequence that 
does not converge.  A similar effect is present in the case of $q=3$ and we have an clique of 4 points (a tetrahedron).

Although we can formally write equations that correspond to extremal directional whitening (i.e. the local equilibrium 
equations), there are graphs such that these equations do not have solution and it is not clear what happens in the 
random case for large $N$.  Something different must be done.

\subsection{Quasi-solutions}

In spite of the fact that the local equilibrium equations may have no solution, we would like to prove that we can 
associate to a local minimum of order $k$ a quasi-solution of the local equilibrium equations.  As we shall see a 
quasi-solution is a function that satisfies the local equilibrium equations nearly everywhere (i.e. in most of the 
nodes, the fraction of 
nodes where the local equilibrium equations are not satisfied goes to zero when $N$ goes to infinity).  In the example 
of previous section is easy to find functions that satisfies equation \ref{TAP.EQ} in all points by one (e.g. 
$c(i,i+1)=1$ for $i$ even and $c(i,i+1)=2$ for $i$ odd).

The proof runs as follows.
Let us consider  a local minimum of order $k$ ($c^{*})$ and a node $i$. The quantity  $\Delta_{b}(c;i)$ is the minimum of 
$H-H^{*}$ when 
the color at $i$ is given by $c(i)$ and the minimum is done over those configurations that differs from $\sigma^{*}$ 
only in nodes at a distance from $i$  
less than $b$ with $b<k$. In the previous notation a node is white if there are al least two different colors such that 
$\Delta_{b}(c;i)=0$.

The quantity $\Delta_{b}(c;i)$ is  a bounded decreasing function of $b$ so that the dependance 
on $b$  is small when $b$ is large. In the case of bicoloring
\be
\Delta_{b}(c;i)=h_{b}(i)(c(i)-c^{*}(i)) \ ,
\ee
where the quantity $h_{b}(i)$ defined by the previous equation is the \emph{effective cavity field}.

In a similar way we can define the quantity $\Delta_{b}(c_{1},c_{2},c_{3};i_{1},i_{2},i_{3})$: it is the minimum of 
configurations that differs from $\sigma^{*}$ only in nodes at a distance from $i_{1}$ or $i_{2}$ or $i_{3}$ less than 
$b$ with $b<k$.

In the case of a random graph we have that 
\be
\Delta_{b}(c_{1},c_{2},c_{3},i_{1},i_{2},i_{3})=\Delta_{b}(c_{1};i_{1})+\Delta_{b}(c_{2};i_{2})+
\Delta_{b}(c_{3};i_{3}) \ , \label{FACTOR}
\ee
with a probability $R(N,\alpha,b)$ such that
\be
\lim_{N \to \infty }R(N,\alpha,b)=1 \ .
\ee
The proof is trivial.  When $N$ goes to infinity at fixed $b$ and $\alpha$ the  set of nodes $N_{k}$ at distance less than 
$b$ from $i_{k}$ has no intersection with probability 1, and the function $\Delta_{b}(c_{1},c_{2},c_{3};i_{1},i_{2},i_{3})$ 
decomposes into the sum of three independent contributions.

We now construct the local equilibrium equations for this problem.  For each site we define the function 
$\Delta_{b}(c;i,j)$, that is defined in the same way as the the function 
$\Delta_{b}(c;i)$, with the only difference that it is computed for the problem where the edge $(i,j)$ is removed.

It is evident that $\Delta_{b}(c;i,j)$ is a function of all $\Delta_{b-1}(c;k,m)$ (for $k\in A(i)$ with $k \ne j$ and 
$m\in A(j)$ with $m \ne i$) with probability 1 (i.e. in all the cases the appropriate factorization equation holds).  The 
explicit expression is rather lengthy and it will be not written here \cite{MP2}.  We can write it
\be
\Delta_{b}(c;i,j)= F(\{\Delta_{b-1}(k,m) \})
\ee
In this way we compute the the quantities $\Delta$ with a value of $b$ using those with $b-1$. 
However in most of the cases for large $b$ the 
functions $\Delta_{b-1}(c;i,j)$ and $\Delta_{b}(c;i,j)$ are the same, so that in the region $1<<b<<N$ the quantities 
$\Delta_{b}(c(k,i))$ satisfies with probability 1 the equations
\be
\Delta_{b}(c;i,j)= F(\{\Delta_{b}(k,m) \})
\ee

These equations are the local equilibrium equations of this problem.  The probability that the factorization equation 
 (\ref{FACTOR}) is  not true at a given node, goes to zero 
with $N$ goes to infinity, however the number of nodes where it is not satisfied may remains constant or go to infinity.  
In this case we are not able to prove that the local equilibrium equations do have a solution, but they have a quasi solution, i.e. 
there are functions $\Delta(c;i,k)$ that satisfy the local equilibrium equations
\be
\Delta(i,j)= F(\{\Delta(k,i) \})
\ee
in most of the sites. More precisely
\be
{\sum_{i,j\in A(i)}|\Delta(i,j)- F(\{\Delta(k,i) \})| \over N}
\ee
should be smaller that an appropriate function of $N$ that goes to zero when $N$ goes to infinity.

In this way we can explicitly associate a quasi-solution of the local equilibrium equations to a $k$-stable configuration. Also in this 
case we expect that many $k$ stable configurations (an whole equilibrium state) do correspond to the same solution of the TAP equations.

\section{Counting the whitening}

The existence of more than one extremal directional whitening corresponds to breaking the phase space into different 
clusters.  This is a very interesting phenomenon.  We can conjecture that the there is an one to one correspondence 
among extremal directional whitenings and clusters of solutions (pure states or lumps).  A natural question is to 
compute the number of extremal directional whitenings that are produced starting by a legal coloring.

On a random graph this can be done analytically using the so called survey propagation equations 
\cite{MP1,MP2,MPZ,PA1,BMZ,QUATTRO}.  For reader 
convenience let us report the main results of \cite{QUATTRO}.  For $\alpha<\alpha_{d}$ there is only one whitening. i. e.  the 
trivial one: all white.  For 
$\alpha_{d}<\alpha<\alpha_{c}$ there is an exponentially large number of solutions, that disappear at $\alpha_{c}$, where 
the graphs are no more colorable.  In the region $\alpha_{d}<\alpha<\alpha_{c}$ the  total number of extremal 
directional whitenings is given by the relation:
\be
\#(extremal \ directional  \ whitening)= \exp(N \Sigma(\alpha)) \ .
  \ee
The complexity $\Sigma(\alpha)$ has been computed. It start from a non-zero value $\Sigma_{d}$ at $\alpha_{d}$, it 
decreases by increasing $\alpha$ and becomes zero  at $\alpha_{c}$. For $\alpha>\alpha_{c}$ directional extremal 
whitenings disappear and a generic graph is no more colorable \cite{QUATTRO,01}.
  
The same techniques we have applied to the local equilibrium equations can be used to study survey propagation 
equations.  The aim of these equations is to characterize the set of all the extremal directional whitenings.  Let us 
define the survey $\vec{\eta}(i)$: the $q+1$ components of this vector are the fraction $\eta_{c}(i)$ of extremal 
whitenings (obtained by whitening a legal coloring) that have the color $c$ at the point $i$.  For symmetry reasons we 
must have that
\bea
\eta_{c}(i)=\eta(i) \ \ \mbox{for} \ \ \  c=1,q \\
\eta_{0}(i)=1 -q \eta(i) \ .
\eea
Let us denote by $\eta_{c_{1},c_{2}} (i_{1},i_{2})$ the fraction of extremal directional whitenings that are obtained by 
whitening a legal coloring that have the color $c_{1}$ at the point $i_{1}$ and the color $c_{2}$ at the point $i_{2}$.  
We can hope that (in analogy with equation \ref{FACTOR}) the following factorization property holds:
\be
\eta_{c_{1},c_{2}} (i_{1},i_{2})=\eta_{c_{1}}(i_{1})\eta_{c_{2}}(i_{1})\ . \
\ee
Such an equation cannot be always exact: it is false if $i_{1}$ and $i_{2}$ are nearby and it is 
 valid only with probability 1 when $N$ goes to infinity. 

Let us define the survey $\vec{\eta}(i,j)$; the $q+1$ components of this vector are the fraction 
$\eta_{c}(i,j)$ of extremal directional whitenings (obtained by whitening a legal coloring) that have the color 
$c$ 
at the point $i$ when the edge from $i$ to $j$ is removed..  

Using the same argument used in the derivation of the 
local equilibrium equations, one can show that 
\be
\vec{\eta}(i,j)= G(\{\vec{\eta}(k,i) \}) \ ,
\ee
where the function $G$ has an explicit form, computed in \cite{QUATTRO}.  The previous equations are the survey 
propagation equations.  

The function $G$ is defined as following: let us call $A_{j}(i)$ the set of adjacent nodes of $i$ that are different 
from $j$ and let as call $d$ its cardinality.  We extract $d-1$ colors, each according to the probability 
$\vec{\eta}(k,m)$ ($k \in A_{j}(i)$) that are interpreted as colors of a directional whitening.  

We call $Z_{j}(i)$ the probability that the number of different non-white colors is less or equal to $q-1$.  The 
quantity $Z_{j}(i)$ is the probability that we can construct a legal directional whitening starting from a whitening 
that does not contains the node $i$ and adding the edges from the node $i$ to the nodes in $A_{j}(i)$.  The quantity 
$\eta_{c}$ are the probabilities (divided by $Z_{j}(i)$) that there are exactly $q-1$ different non-white colors and 
that the missing color is $c$.

Using the solution of the previous equations, that is supposed to be unique \cite{PA1} on a random graph for large $N$, we can 
estimate the complexity density \cite{MPZ,MZ}, using
\be
N \Sigma = \sum_{i} \ln(Z(i)) - \frac12 \sum_{i,j\in A(i) }\ln (Z(i,j)) \,
\ee
where $Z(i,j)$ is the probability that, when we extract a color in $i$ with probability $\vec{\eta}(i,j)$ and we extract a 
color in $j$ with probability $\vec{\eta}(j,i)$,  the two colors are different.  In other words $Z(i,j)$ is the 
probability to be able to produce a legal directional whitening if we add the edge from $i$ to $j$ to a graph that does 
not contains it.  In the same way $Z(i)$ is the probability to be able to produce a legal  whitening if we 
add the node $i$ to a graph that does not contains it and is is given by 
\be
Z(i)=Z_{j}(i)Z(i,j)
\ee
and it does not depend on the choice of of $j \in A(i)$.

The arguments leading to the previous equations will be not reported here and they can be found in the original papers.

\section{Conclusions}

In the colorable phase for coloring (or in the satisfiable phase for K-satisfiability) we have shown that 
one can associate solutions of local equilibrium equations to extreme directional whitenings that are generated starting 
from legal configurations.  We also conjecture that a cluster of solutions correspond to one extreme directional whitening.  
If the legal configurations cluster in many different states, the existence of many extremal directional whitenings is a 
signal of the existence of many states.  If we identify, as supported by these considerations,  extremal directional 
whitenings with local 
equilibrium states, the existing predictions on the number (and on the other properties) of local equilibrium states, 
\cite{MPZ,QUATTRO} can be tested by studying the extremal directional whitening.  The construction and the counting of 
the extremal directional whitening is numerically an easy job, so that the prediction of this note and of the whole 
replica based approach \cite{MP1,MP2} can be tested numerically without difficulties.

An other important point is the proof that the local equilibrium equations (e.g. TAP equations or belief equations) in 
the uncolorable phase must be satisfied nearly everywhere, but no argument can be done that implies that the equations 
should be satisfied everywhere: quasi solution must exist but \emph{bona fide} solutions may be missing.

%\end{Large}
\end{document}